\documentclass[multphys,vecphys]{svmult}

\usepackage{makeidx}     
\usepackage{graphicx}    
\usepackage{multicol}    

\makeindex             

\begin{document}

\title*{Quantum $1/f$ Noise in Equilibrium: \\from Planck to Ramanujan}
\author{Michel Planat}
\institute{Laboratoire de Physique et M\'etrologie des
Oscillateurs du CNRS, \\32 Avenue de l'Observatoire, 25044 Besan\c
con Cedex, France.
}
%
%
\maketitle

\begin{abstract}
We describe a new model of massless thermal bosons which predicts
an hyperbolic fluctuation spectrum at low frequencies. It is found
that the partition function per mode is the Euler generating
function for unrestricted partitions $p(n)$. Thermodynamical
quantities carry a strong arithmetical structure: they are given
by series with Fourier coefficients equal to summatory functions
$\sigma_k(n)$ of the power of divisors, with $k=-1$ for the free
energy, $k=0$ for the number of particles and $k=1$ for the
internal energy. Low frequency contributions are calculated using
Mellin transform methods. In particular the internal energy per
mode diverges as $\frac{\tilde{E}}{kT}=\frac{\pi^2}{6 x}$ with
$x=\frac{h \nu}{kT}$ in contrast to the Planck energy
$\tilde{E}=kT$.  The theory is applied to calculate corrections in
black body radiation and in the Debye solid. Fractional energy
fluctuations are found to show a $1/\nu$ power spectrum in the low
frequency range. A satisfactory model of frequency fluctuations in
a quartz crystal resonator follows. A sketch of the whole
Ramanujan--Rademacher theory of partitions is reminded as well.

\end{abstract}

\section{Introduction}               
According to the equipartition law of statistical mechanics, the
available noise power $\tilde{P}$ in the frequency interval $d\nu$
is equal to $kTd\nu$ \cite{VanderZiel}: this result is essentially
Nyquist's theorem for the voltage noise $ \left < v^2 \right >$ at
a resistor $R$, i.e. $\tilde{P}=\frac{\left <v^2 \right >}{4R}=kT
d\nu $, where $\left < \right
> $ means the average value. Since $\tilde{E}=kT$ is the mean energy per
mode, Nyquist proposed to add quantum corrections as
$\frac{\tilde{E}}{kT}=p(x)$, with the Planck's factor
$p(x)=\frac{x}{\exp(x)-1}$ in which $x=h\nu/kT$. This result was
generalized as $\frac{\tilde{E}}{kT}=p(x)+\frac{x}{2}$ to account
for the zero point energy. There are still controversies
concerning the physical relevance of these relations: the Planck
factor removed the ultaviolet divergence but this was
reincorporated in the frame of quantum electrodynamics \cite
{Abbott},\cite{Callen}.

At the present stage, quantum statistical mechanics does not
include infrared corrections of the $1/\nu$ type. The infrared
catastrophe was studied previously in non-stationary processes
such as the scattering of electrons in an atomic field
\cite{Bloch},\cite{Handel0}.

Here we derive an alternative approach in which $1/\nu$ noise is a
property of non degenerate bosons in equilibrium. We first observe
that the quantum mechanical partition function of a boson gas,
with equally spaced energy levels, is the Euler generating
function of $p(n)$: the number of indiscernible collections of the
integer $n$. It relates to elliptic modular functions which are
very exactly known. As a result the main contribution in the mean
energy per mode is the infrared term
$\frac{\tilde{E}}{kT}=\frac{\pi^2}{6x}$ instead of unity. The new
law also leads to $1/\nu$ fractional energy fluctuations of the
whole gas.

Using the new approach and the density of states of the
conventional approach we calculate the corrections to black-body
radiation laws, including the density of photons, the emissivity
and infrared fluctuations. We also apply the calculations to the
phonon gas in a quartz resonator.

The partition function $Z$ of a non degenerate boson gas is given
from
\begin{equation}
\ln Z=-\sum_s \ln[1-\exp(-\beta \epsilon_s)], \label{equa1}
\end{equation}
where the summation is performed over all the states $s$ of the
assembly. In the conventional approach it is thus considered that
the partition function $\tilde{Z}$ per mode of frequency
$\epsilon_s=h \nu_s$ is such that $\ln
\tilde{Z}=-\ln[1-\exp(-\beta\epsilon_s)]$.

In black-body radiation one accounts for the wave character of the
quantum states by counting the number $l$ of wavelenghts in a
cubic box of size $L$
\begin{equation}
l^2=\frac{\nu_s^2 L^2}{c^2}=l_1^2+l_2^2+l_3^2, \label{equa2}
\end{equation}
where the summation (\ref{equa1}) should be performed over all
integers $l_1,~l_2,~l_3$ obeying  (\ref{equa2}).

This can be achieved by removing the discretness of energy levels
and replacing the sum (\ref{equa1}) by an integral
\begin{equation}
\ln Z=-\int_0^{+\infty}D(\nu) \ln[1-\exp(-\beta h\nu)]d\nu,
\label{equa3}
\end{equation}
with $D(\nu)=2 \times \frac{4 \pi V \nu^2}{c^3}$ the density of
states \cite{Kestin}: the factor $2$ happens due to the two
degrees of freedom of polarization, $c$ is the light velocity and
$V=L^3$ the volume of the cavity.

From now we consider that to each mode is associated a set of
equally spaced energy levels $n h \nu$, $n$ integer, so that the
partition per mode becomes
\begin{equation}
\ln \tilde{Z}=-\sum_{n\ge 1} \ln[1-\exp(-n \beta h \nu].
\label{equa4}
\end{equation}
As shown in Sect. (\ref{Euler}) this accounts for new
multiparticle microstates not considered so far.

The summation above is well known in number theory and can be very
accurately described using elliptic modular functions. At is will
be shown, there are drastic consequences in the low frequency part
of the spectrum, while the high frequency part is left unchanged.

In the following the thermodynamical quantities will be defined as
usual
\begin{eqnarray}
&& N=-\frac{\partial \ln Z}{\partial(\beta\epsilon_s)},
~~~~~~\rm{the~occupation~ number},\\ && E=- \frac{\partial \ln
Z}{\partial
\beta},~~~~~~~\rm{the~internal~energy},\\
&& S=
\frac{\partial F}{\partial T},~~~~~~~~~~~~~\rm{the~entropy},\\
&&u=kT^2\frac{\partial \ln Z}{\partial T},  ~~~~~\rm{the~spectral~
energy~ density}, \\&&F=-kT\ln Z,  ~~~~~~\rm{the~free~energy}, \\
&& \epsilon^2 =kT^2\frac{\partial E}{\partial T},~~~~~~~\rm{
the~fluctuations~ ~ of~ the~ internal~ energy}. \label{fluctu}
\end{eqnarray}
In all the paper the subscript $^{\sim}$ will indicate that we
restrict the calculation to one single mode.

\section{Thermodynamics of the Euler Gas}
\label{Thermo}
\subsection{Euler generating function}
\label{Euler} The partition function per mode $\tilde Z$ in
(\ref{equa4}) can be written in the Euler form \cite{Rademacher}
\begin{equation}
\tilde{Z}(y)=\prod_{n\ge 1}\frac{1}{1-y^n}=\sum_{n\ge 1}p(n)y^n,
\label{equa5}
\end{equation}
with $y=\exp (-x)$ and $x=\frac{h \nu}{k T}$. This is equivalent
to the Boltzmann summation
\begin{equation}
\tilde{Z}(y)=\sum_{n\ge 1}p(n) \exp(-nx), \label{equa6}
\end{equation}
where $p(n)$ is the degeneracy parameter of the energy level
$nh\nu$. It is known in number theory as the number of
unrestricted partitions of the integer $n$, that is the number of
different ways of calculating $n$ as  a sum of integers.

For example with $n=4$ we have $p(4)=5$ and the corresponding
indiscernible collections are
\begin{eqnarray}
&& 4=4+0+0+0~~~~~~(a), \nonumber\\
&&4=1+1+1+1~~~~~~(b),~~~~4=2+2+0+0~~~~~~(c), \nonumber \\
&&4=3+1+0+0~~~~~~(d),~~~~4=2+1+1+0~~~~~~(e). \nonumber \\
\label{colle}
\end{eqnarray}
This can be pictured in terms of the energy levels. The collection
(a) means one particle on the level of index $4$ and the three
remaining particles on the ground state of index $0$, i.e. $4
h\nu=1 \times 4 h \nu+3\times 0 h \nu$. This collection is the
only one considered in the conventional approach. The others
microstates from (b) to (d) corresponds to different possibilities
of bunching of the particles. Collection (b) means the four
particles on the level of index $1$, i.e. $4 h \nu=4\times h \nu$,
collection (c) means two particles on the level $2$ and two
particles on the ground state, i.e. $4h\nu=2\times2h\nu+2\times 0
h\nu$, and so on.

Properties of Euler generating function were studied in full
details by Ramanujan \cite{Ramanujan} in 1918 and completed by
Rademacher \cite{Rademacher} in 1973. An important result is the
asymptotic formula
\begin{equation}
p(n)\sim
\frac{1}{4n\sqrt{3}}\exp(\pi\sqrt{2n/3})~~\rm{when}~n\rightarrow
\infty. \label{equa7}
\end{equation}
\begin{figure}[htbp]
\centering{\resizebox{10cm}{!}{\includegraphics{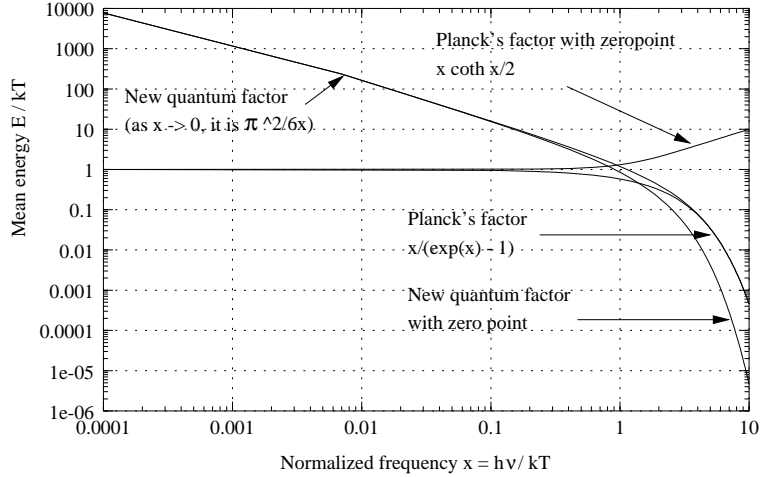}}}
\caption{Comparison of the free energy from the conventional
Planck's theory and from the quantum $1/\nu$ theory.}
\end{figure}
If instead of eq. (\ref{equa5}) one uses the classical partition
function \\ $Z=\sum_{n\ge 1}\exp(-nh\nu/kT)$, one recovers the
Planck's factor $E/kT=x/(\exp(x)-1)$. Such a dependence remains
valid at high frequencies $h\nu»kT$. If one introduces the zero
point energy through the formula
$\epsilon_{s,n}=(n+\frac{1}{2})h\nu$ with $n\ge 1$ instead of
$\epsilon_{s,n}=n h\nu$, then the low frequency part of the
quantum spectrum is left unchanged but the ultraviolet catastrophe
that one gets from the formula $E/kT=x/(\exp(x)-1)+x/2=x\coth
(x/2)$, is cancelled as shown in Fig. 1. The new quantum theory
thus solves simultaneously the questions concerning the $1/\nu$
power spectrum (which is observed but was not predicted) and the
ultraviolet catastrophe (which is not observed but was predicted)
\cite{Callen}.
\newpage
\subsection{Riemann zeta function and the free energy \\ of the Euler gas}
\label{Riemann} The partition function $\tilde{Z}(x)$ defined in
(\ref{equa4}) is related to the Riemann zeta function $\zeta(s)$
through the Mellin transform as follows (Ref. \cite{Ninham}, Eq.
(6.3) p. 464)
\begin{equation}
\Gamma(s) \zeta(s)\zeta(s+1)=\int_0^{\infty} (-\ln \tilde{Z}(x))
x^{s-1} dx. \label{equa8}
\end{equation}
Introducing $\sigma_k(n)$ as the sum of $k^{\rm{th}}$ powers of
the divisors of $n$ and the Dirichlet series
\begin{equation}
\zeta(s)\zeta(s-k)=\sum_{n\ge 1}\frac{\sigma_k (n)}{n^s},
\label{equa9}
\end{equation}
and computing the inverse Mellin transform one obtains
(\cite{Ninham}, p. 467) the free energy $\tilde{F}$ as follows
\begin{equation}
\frac{\tilde{F}}{kT}=-\ln \tilde{Z}(x)=\sum_{n\ge
1}\ln(1-\exp(-nx)) =-\sum_{n\ge 1} \sigma_{-1}(n) \exp(-nx).
\label{equa10}
\end{equation}
with $\sigma_{-1}(n)=\sigma_{1}(n)/n$. There is thus a close
relationship between the arithmetic of $p(n)$ and that of
divisors. This will be confirmed in the derivation of the others
thermodynamical quantities.

Since the main contributions to $\tilde{Z}(x)$ are given by the
poles at $s=0$ of $\Gamma(s)$ and $\zeta(s+1)$ and at  $s=1$ of
$\zeta(s)$, the free energy may be approximated in the low
frequency part of the spectrum (\cite{Elizalde}, p. 58)
\begin{equation}
\frac{\tilde{F}}{kT} \simeq -\frac{\pi^2}{6 x}-\frac{1}{2}\ln
(\frac{x}{2 \pi})+ \frac{x}{24}, \label{equa11}
\end{equation}
with the error term $\sum_{l\ge 1}\ln(1-\exp(-4\pi^2l/x))$.

\subsection{Dedekind eta function and the internal energy of the Euler gas}
One easily shows that the Mellin transform of the occupation
number $\tilde{N}(x)$ is $\Gamma(s)\zeta(s)^2$. A similar
derivation to the one performed in Sect. \ref{Riemann} leads to
\begin{equation}
\tilde{N}(x)=-\frac{\partial (\ln \tilde{Z}(x))}{\partial( n
x)}=\sum_{n\ge 1}\frac{1}{\exp(nx)-1}=\sum_{n\ge 1}
\sigma_0(n)\exp(-n x). \label{equa12}
\end{equation}
In the low frequency region one gets (see \cite{Flajolet}, p. 27)
\begin{equation}
\tilde{N}(x) \simeq \frac{-\ln x + \gamma}{x}. \label{equa13}
\end{equation}
where $\gamma(x)\simeq 0.577$ is Euler constant.

The Mellin transform of the internal energy $\tilde{E}(x)$ is
$\Gamma(s)\zeta(s)\zeta(s-1)$ and from the same method than above
\begin{equation}
\tilde{E}(x)=-\frac{\partial (\ln \tilde{Z}(x))}{\partial (\beta)}
=h\nu \sum _{n\ge 1}\frac{n}{\exp(nx)-1}=h\nu \sum_{n\ge 1}
\sigma_1(n)\exp(-n x). \label{equa14}
\end{equation}
An alternative derivation involves elliptic modular aspects.
According to Ninham \cite{Ninham}: \textit{All mathematics is a
tautology, and all physics uses mathematics to look in different
ways at a fundamental problem of philosophy - how to bridge the
discrete and continuous}.

The link between the modular group and the Euler generating
function is from the equality \cite{Rademacher}
\begin{equation}
\tilde{Z}(y)=\prod_{n\ge 1}\frac{1}{1-y^n}=\sum_{n\ge
1}p(n)y^n=\frac{\exp{(i\pi\tau/12)}}{\eta(\tau)},
\label{equa14bis}
\end{equation}
where the domain of integration of $\tilde{Z}(y)$ is taken to be
the upper half complex plane of the new variable $\tau$
\begin{equation}
y=\exp(2 i \pi \tau),~~\Im(\tau)>0. \label{equa15}
\end{equation}
Here we have
\begin{equation}
\Im(\tau)=\frac{x}{2 \pi},~~x=\frac{h \nu}{kT}. \label{equa16}
\end{equation}
As shown in Sect. (\ref{Mathematical}) Dedekind eta function acts
on the full modular group $SL(2,Z)$. At this stage we do not enter
into the full ramifications of the theory and only emphasizes the
connexion to the modular Eisenstein function
\begin{equation}
G_2(\tau)=\sum'_{m,n} \frac{1}{(m \tau +n)^2}, \label{equa18}
\end{equation}
where the summation is performed over all non zero relative
integers $m$ and $n$ and $\Im(\tau)>0$. It can be shown
(\cite{Weil}, p. 29) that $G_2(\tau)$ connects to the logarithmic
derivative of $\eta(\tau)$
\begin{equation}
G_2(\tau)=-4 i \pi \frac{d(\ln (\eta(\tau)))}{d \tau},
\label{equa19}
\end{equation}
with a Fourier expansion
\begin{equation}
G_2(\tau)=2 \zeta(2)+2(2i\pi)^2\sum_{n\ge 1}\sigma_1(n)exp(2 i \pi
n \tau). \label{equa20}
\end{equation}
Using (\ref{equa14bis}), (\ref{equa15}) and
(\ref{equa18})-(\ref{equa20}) the relation  (\ref{equa14}) is
easily recovered.

The low frequency expansion of internal energy is as follows
\begin{equation}
\frac{\tilde{E}}{kT}\simeq \frac{\pi^2}{6
x}-\frac{1}{2}+\frac{x}{24}, \label{equa21}
\end{equation}
instead of the Planck result $\tilde{E}\simeq kT$.

One can also compute the entropy $\tilde{S}$ from
\begin{eqnarray}
&&\frac{\tilde{S}}{k}=\frac{\tilde{E}}{k T}+
\ln(\tilde{Z})=\frac{\tilde{E}-\tilde{F}}{kT}=\sum_{n\ge 1}
\sigma_1(n)\left(x+1/n\right)
\exp(-n x)\nonumber \\
&&\simeq \frac{\pi^2}{3 x}+\frac{1}{2}\ln \frac{x}{2
\pi}-\frac{1}{2}~~\rm{when}~~x \rightarrow~0. \label{equa22}
\end{eqnarray}

At very low frequency it results that the internal energy equals
the opposite of free energy and one half the entropy.

\section{Application to Black-body Radiation}
\subsection{Stefan-Boltzmann constant revisited}
\label{emissivity} The Stefan-Boltzmann constant is an integrated
measure of the emissivity of a black body \cite{Kestin}. In the
conventional approach the partition function is calculated from
the integral (\ref{equa3}) using the density of states $D(\nu)=
\frac{8 \pi V \nu^2}{c^3}$ that is
\begin{equation}
\ln Z=8\pi V\left(\frac{kT}{ch}\right)^3\times2\zeta(4),
\label{equa23}
\end{equation}
with $\zeta(4)=\pi^4/90$ and we used the Mellin integral formula
\begin{equation}
\zeta(s+1)=-\frac{1}{\Gamma(s)}\int_0^{\infty}x^{s-1}\ln(1-\exp(-x))dx.
\label{equa24}
\end{equation}
The Stefan-Boltzmann constant $\sigma_{\rm{SB}}$ is defined from
the free energy
\begin{equation}
F=-kT\ln
Z=-\frac{4\sigma_{\rm{SB}}}{3c}VT^4~~\rm{with~}\sigma_{\rm{SB}}=
\frac{2\pi^5k^4}{15c^2 h^3}. \label{equa23bis}
\end{equation}

If instead of (\ref{equa23}) one uses the general formula
\begin{equation}
\ln Z=-8\pi V\left(\frac{kT}{ch}\right)^3 \int_0^{\infty}
x^2\sum_{n\ge1}\ln (1-exp(-n x)) dx, \label{equa23ter}
\end{equation}
the interchange of the integral and the sum leads to
\begin{equation}
\ln Z=-8\pi V\left(\frac{kT}{ch}\right)^3 \times 2 \zeta(4) \times
\left( \sum_{n\ge 1}\frac{1}{n^3}\right). \label{equa23quatro}
\end{equation}
As a result we find a free energy (and a modified Stefan-Boltzmann
constant) in excess with a factor $\sum_{n\ge 1}\frac{1}{n^3}
=\zeta(3)\simeq 1.20$. If one uses the alternative derivation in
terms of the divisors one recovers  the mathematical formula
(\ref{equa9}) with $s=3$ and $k=-1$.

\subsection{The density of photons}
The number of photons in the bandwidth $d\nu$ is
\begin{equation}
dN(\nu,T)=D(\nu) \tilde{N}(\nu,T) d\nu, \label{equa24bis}
\end{equation}
with the occupation number $\tilde{N}(\nu,T)=(\exp(\beta h
\nu)-1)^{-1}$ in the conventional approach. Integrating one gets
per unit volume
\begin{equation}
\frac{N(T)}{V}=8\pi \left(\frac{ kT}{ ch} \right)^3
\int_0^{+\infty}\frac{x^2 dx}{\exp(x)-1}=8\pi
\left(\frac{kT}{ch}\right)^3 \times 2\zeta(3). \label{equa25}
\end{equation}

In the general approach the occupation number is defined from the
summation (\ref{equa12}) and we need to evaluate
\begin{equation}
 \int_0^{\infty}x^2 \sum
_{n\ge1}\frac{1}{\exp(nx)-1} =2 \zeta(3)\left( \sum_{n\ge
1}n^{-3}\right)=2 \zeta(3)^2. \label{equa26}
\end{equation}
This is in excess of a factor $\zeta(3)\simeq 1.20$ as for the
free energy. If one compares the calculation in terms of divisors
one recovers the mathematical formula (\ref{equa9}) with $s=3$ and
$k=0$.

\subsection{Planck's radiation formula revisited}
The energy within the bandwidth $d\nu$ is defined as
\begin{equation}
dE(\nu,T)=D(\nu) \tilde{E}(\nu,T) d\nu=u(\nu,T)d \nu,
\label{equa27}
\end{equation}
with $\tilde{E}(\nu,T)=h\nu (\exp(\beta h \nu)-1)^{-1}$ in the
conventional approach and with $u(\nu,T)$ the energy spectral
density. We get the Planck's radiation formula
\begin{equation}
u(\nu,T)=\frac{8\pi hV}{c^3}\frac{\nu^3}{\exp(\beta h \nu)-1}.
\label{equa28}
\end{equation}
The black-body emissivity is defined as
\begin{equation}
e_b(\nu,T)=\frac{c}{4V}u(\nu,T). \label{equa29}
\end{equation}
At very low frequency the conventional result leads to the
Rayleigh-Jeans formula
\begin{equation}
\left [e_b(\nu,T)\right]_{\rm{RJ}}=2\pi\frac{k}{c^2}\nu^2 T,
\label{equa30}
\end{equation}
which is independent of Planck's constant and is proportional to
the inverse of the square of wavelength $\lambda=c/\nu$.

In the new approach the emissivity is
\begin{equation}
e_b(\nu,T)=\frac{2 \pi h}{c^2}\sum_{n\ge
1}\frac{n\nu^3}{\exp(n\beta h \nu )-1}. \label{equa31}
\end{equation}
At very low frequency one uses (\ref{equa21}) with the result
\begin{equation}
\left [e_b(\nu,T)\right]_{\rm{LF}}=\frac{\pi^3}{3}\frac{k^2}{c^2
h}\nu T^2. \label{equa33}
\end{equation}
Thus the $\nu^2 T$ dependance is replaced by the $\nu T^2$
dependance, the low frequency emissivity now depends on the Planck
constant and on the inverse wavelength; there is a ratio
$\frac{\pi^2}{6 x}$ between the new result and the one predicted
by the Rayleigh-Jeans formula.

\subsection{Radiative atomic transitions}
Let us now consider the equilibrium between atoms and a radiation
field, allowing the emission or absorption of photons of frequency
\begin{equation}
\nu=\nu_1-\nu_2, \label{spont_1}
\end{equation}
where $\epsilon_2=h\nu_2$ is the energy in the upper state and
$\epsilon_1=h\nu_1$ in the lower state.

The conventional theory, as derived for the first time by
Einstein, states that the rate at which atoms make a transition
$1\rightarrow 2$ in which one photon is absorbed is equal to the
rate at which atoms emit photons, so that
\begin{equation}
B_{12}N_1 u(\nu)=A_{21}N_2+B_{21}N_2 u(\nu), \label{spont_2}
\end{equation}
where $N_1$ and $N_2$ are the occupation numbers of atoms in
levels $1$ and $2$, $A_{21}$ is the spontaneous absorption rate,
$B_{12}$ is the induced emission rate and $u(\nu)$ is the energy
density in the radiation field as given in (\ref{equa28}).

In thermal equilibrium the occupation numbers in states $1$ and
$2$ obey the Boltzmann law
\begin{equation}
\frac{N_2}{N_1}=\exp(-\frac{h\nu}{kT}). \label{spont_3}
\end{equation}
Using (\ref{spont_3}) and (\ref{equa28}) one gets the well known
formulas
\begin{equation}
A_{21}=A,~~B_{12}=B_{21}=B~~\rm{and}~~\frac{A}{B}=\frac{8\pi
h}{\lambda^3}, \label{spont_4}
\end{equation}
where $\lambda=c/\nu$ is the wavelength of the radiation field.

If one uses the general formula one gets low frequency corrections
in the spontaneous to stimulated emission ratio. Using the low
frequency expression (\ref{equa21}) for the internal energy this
yields
\begin{equation}
\left[\frac{A}{B}\right]_{\rm{LF}}=\frac{A}{B}\times
\frac{\pi^2}{6 x} =\frac{4\pi^3 kT}{3c\lambda^2}. \label{spont_5}
\end{equation}
The $A/B$ low frequency ratio now depends on the inverse of the
square of wavelength $\lambda$ and is independant on the Planck
constant $h$, in contrast to the $h/\lambda^3$ dependance of the
standard ratio. The spontaneous to stimulated absorption rate is
enhanced a factor $\frac{\pi^2}{6 x}$ over the conventional one.

\subsection{Einstein's fluctuation law revisited}

According to the conventional Einstein's approach \cite{Pais} the
energy fluctuations of a system in equilibrium within a larger
system of temperature $T$ are
\begin{equation}
 \epsilon^2=\left <(E-\left< E\right>)^2 \right >=-\frac{\partial \left<
 E\right>}{\partial \beta}
=kT^2\frac{\partial \left< E\right>}{\partial T}. \label{equa34}
\end{equation}
One can reformulate this relation for the fluctuations of the
energy $dE=u(\nu,T) d\nu$ in the bandwidth $d\nu$
\begin{equation}
d \epsilon^2 =k T^2\frac{\partial u}{\partial T} d\nu =S_u(\nu)
d\nu, \label{equa34bis}
\end{equation}
with $u$ the energy spectral density  and $S_u(\nu)$ the power
spectral density of the fluctuations of $u$. One gets
(\cite{Pais}, p. 429 )
\begin{equation}
 d \epsilon^2=(h\nu u+\frac{c^3}{8\pi\nu^2 V}u^2)Vd\nu.
\label{equa35}
\end{equation}
The first term at the right hand side is the one corresponding to
the high frequency part of the spectrum (Wien's law): it is of a
pure quantum nature and is corpuscular like; the second one
corresponds to the low frequency (Rayleigh-Jeans) region: it is
purely classical and wavelike. In the low frequency part of the
spectrum they are fractional energy fluctuations of the random
walk type
\begin{equation}
\left[\frac{S_u(\nu)}{u^2}\right]_{\rm{ RJ}}=\frac{c^3}{8\pi
V}\frac{1}{\nu^2}. \label{equa36}
\end{equation}

In the new approach one uses the low frequency energy density
$u(\nu,T)\simeq \frac{4\pi^3 V}{3 c^3 h}\frac{\nu d\nu}{\beta^2}$
so that instead of (\ref{equa36}) one gets
\begin{equation}
\left[\frac{S_u(\nu)}{u^2}\right]_{\rm{LF}}
=\frac{3}{2}\frac{hc^3}{\pi^3 V}\frac{1}{kT\nu}.
\label{Einstein_5}
\end{equation}
This is the announced quantum $1/\nu$ fluctuation spectrum. There
is a reduced low frequency noise and the ratio between the new
result and the Einstein-Rayleigh-Jeans  one is $\frac{12
x}{\pi^2}$.

\section{Application to a Phonon Gas and to the $1/f$ Frequency Noise of a Quartz Resonator}
\subsection{The specific heat of a phonon gas revisited}

The properties of the phonon gas are quite similar to those of the
photon gas except for the new form of the density of modes as
$g(\nu)=\frac{12 \pi V}{c_{\rm{ph}}^3}\nu^2$ with
$\frac{3}{c_{ph}^3}=\frac{2}{c_t^3}+\frac{1}{c_l^3}$ where
$c_{\rm{ph}}$ represents the average wave velocity and $c_t$ and
$c_l$ are the transverse and longitudinal velocities for an
isotropic solid \cite{Kestin}. The maximal vibrational frequency
$\nu_m$ (Debye frequency) is defined from the total number of
allowed quantum states
\begin{equation}
3N_0=\frac{12 \pi V}{c_{\rm{ph}}^3}
\int_0^{\nu_{\rm{m}}}\nu^2d\nu~~\rm{that~is}~\nu_m= \left(
\frac{3N_0c_{\rm{ph}}^3}{4\pi V}\right)^{1/3}, \label{phonon_1}
\end{equation}
where $N_0$ is the number of atoms in the volume $V$.

In the conventional theory \cite{Kestin} we get
\begin{equation}
\ln Z_{\rm{ph}}=-\frac{9N_0}{\nu_{\rm{m}}^3}
\int_0^{v_{\rm{m}}}\nu^2 \ln\left[ 1-\exp(-\beta h \nu)\right]
d\nu. \label{phonon_2}
\end{equation}
The internal energy follows from the formula
\begin{equation}
E=\frac{9RT}{x_{\rm{m}}^3}D(x_{\rm{m}}), \label{phonon_3}
\end{equation}
and the constant volume specific heat $C_{\rm{v}}=\partial
E/\partial T$ equals
\begin{equation}
C_{\rm{v}}=3R\left[ D(x_{\rm {m}})-x_{\rm{m}}
D'(x_{\rm{m}})\right], \label{phonon_4}
\end{equation}
with $D(x_{\rm {m}})=\frac{3}{x_{\rm {m}}^3}\int_0^{x_{\rm
{m}}}\frac{x^3 dx}{\exp(x)-1}$  the Debye function, and
$x_{\rm{m}}=\frac{\theta_{\rm{D}}}{T}$, with
$\theta_{\rm{D}}=h\nu_{\rm{m}}/k$ the Debye characteristic
temperature.

The case $T»\theta_{\rm{D}}$, $D(\theta_{\rm{D}})\rightarrow 1$
corresponds to the Dulong-Petit value $C_{\rm{v}}\sim3R$. At very
low temperatures one gets the cubic temperature dependence
$C_{\rm{v}}\sim \frac{4\pi^4}{5}\times3R\left (
\frac{T}{\theta_{\rm{D}}}\right)^3$.

In the new approach
\begin{equation}
\ln Z_{\rm{ph}}=-\frac{9N_0}{\nu_{\rm{m}}^3}
\int_0^{v_{\rm{m}}}\sum_{n\ge 1}\nu^2 \ln\left[ 1-\exp(-n\beta h
\nu)\right] d\nu. \label{phonon_5}
\end{equation}
Debye results are found unchanged except for an extra
multiplicative factor in the specific heat as was the case for the
integrated emissivity in Sect. \ref{emissivity} that is
\begin{equation}
\frac{\left[ C_{\rm{v}} \right]_{\rm{new}}}{C_{\rm{v}}
}=\zeta(3)\sim 1.20. \label{phonon_6}
\end{equation}
At very low temperatures the electronic contribution to the
specific heat which decreases as $T$, dominates the lattice
contribution, which decreases as $T^3$.  This is accounted for in
the conventional way.

\subsection{1/f noise in a quartz resonator}

Specific heat is involved in the energy fluctuations of a
canonical ensemble from the relation
\begin{equation}
\epsilon^2=kT^2 C_{\rm{v}}. \label{phonon_7}
\end{equation}
The relative energy fluctuations follows as
$\frac{\epsilon^2}{E^2}=\left(\frac{2}{3N_0}\right)^{1/2}$ which
is on the order $10^{-11}$ for $N_0=10^{23}$, the Avogadro number.

For energy fluctuations in the bandwidth $d\nu$, the main
difference with the conventional theory lies in the low frequency
region, as was the case of the photon gas. We find the quantum
$1/\nu$ formula
\begin{equation}
\left[\frac{S_u(\nu)}{u^2}\right]_{\rm{LF}}
=\frac{9hc_{\rm{ph}}^3}{4\pi^3
V}\frac{1}{kT\nu}=\frac{A_{\rm{ph}}}{V\nu}. \label{phonon_8}
\end{equation}
The method can be used to predict fractional frequency
fluctuations in a quartz crystal resonator from the formula
\footnote{To establish the formula one writes the equation for a
lossy harmonic oscillator and one postulates that the $1/\nu$
fluctuations are present in the loss coefficient. }
\cite{Gagnepain}-\cite{Handel}
\begin{equation}
\frac{S_{\omega}(\nu)}{\omega^2}=\frac{1}{4Q^4}\frac{A_{\rm{ph}}}{V\nu}=\frac{h_{-1}}{\nu}.
\label{phonon_9}
\end{equation}
where $\omega$ and $Q$ are the frequency and quality factor of the
resonator. Using $c_{\rm{ph}}\sim3.5\times10^3~m/s$, we find
$A_{\rm{ph}}\sim 5\times10^{-4}$. For a $5$ MHz P5 quartz crystal
resonator with $Q\sim 2\times10^6$, the active region under the
electrodes has thickness $t=5\lambda/2\sim 3~mm$, and section
$S\sim3~cm^2$, that is $V\sim 1~cm^3$. The resulting $1/\nu$
factor is $h_{-1}=\frac{A_{\rm{ph}}}{4Q^4V}\sim
\frac{10^2}{Q^4}\sim6\times10^{-24}$. This is the order of
magnitude found in experiments \cite{Gagnepain}.

\section{Ramanujan--Rademacher Theory of Partitions:\\ a Short Reminder }
\label{Mathematical}

Besides the low frequency approximations encoutered in Sect.
(\ref{Thermo}) there is a an exact method to calculate the number
of partitions $p(n)$ first discovered by Ramanujan
\cite{Ramanujan} and improved by Rademacher \cite{Rademacher}
thanks to an integration along Ford circles in the complex half
plane. For completness we remind here the main points of the
theory from which the results in Sect. (\ref{Thermo}) may also be
derived .

From well-known mathematical arguments \cite{Ramanujan} (p. 113)
one can get the leading term for the case $0<y<1$ and
$y\rightarrow 1$ from the expression \samepage\footnote{We have
\begin{eqnarray}
\log Z(y)=\sum_n \log
\frac{1}{1-y^n}=\sum_{m,n}\frac{y^{mn}}{m}=\sum _m
\frac{y^m}{m(1-y^m)}\nonumber ,
\end{eqnarray}
and
\begin{eqnarray}
my^{m-1}(1-y)<1-y^m<m(1-y)\nonumber ,
\end{eqnarray}
so that
\begin{eqnarray}
\frac{1}{1-y}\sum_m \frac{y^m}{m^2}<\log Z(y) <\frac{1}{1-y}\sum_m
\frac{y}{m^2}\nonumber
\end{eqnarray}
Each of the above series has the limit $\pi^2/6$ when $y
\rightarrow 1$ and so $\log Z(y)\sim \frac{\pi^2}{6(1-y)}$}.
%

\begin{equation}
\ln \tilde{Z}(y)\sim \frac{\pi^2}{6(1-y)}. \label{logZ_3}
\end{equation}
The use of $y=exp(-h\nu/kT)$ corresponds to the low frequency
approximation at $\nu \rightarrow 0$, that is
$1-y=1-\exp(-h\nu/kT)\sim h\nu/kT$. This leads to the leading low
frequency term in the free energy (\ref{equa11}) and internal
energy  (\ref{equa21}).

They are similar formulas associated with rational points which
are located at
\begin{equation}
y_{\rm{pq}}=\exp(2i\pi\frac{p}{q}), \label{math_01}
\end{equation}
on the unit circle $|y|=1$. The leading term in the expansion of
$\tilde{Z}(y)$ corresponds to the fundamental mode
$\frac{p}{q}=\frac{1}{1}$.

The general method to compute rational contributions is a master
piece of twentieth century mathematics (\cite{Ramanujan}),
(\cite{Rademacher}). It uses the connexion of $\tilde{Z}(y)$ to
the elliptic modular functions.

\subsection{The fundamental contribution }
To compute the contribution of the fundamental point $1/1$ of the
unit circle $|y|=1$ one uses the property (\cite{Apostol}, p. 96,
\cite{Elizalde}, P. 58)
\begin{eqnarray}
&&\tilde{Z}(y)=\frac{y^{1/24}}{\sqrt{2\pi}}(\ln\frac{1}{y})^{1/2}\exp\left[
\frac{\pi^2}{6\ln \frac{1}{y}}\right]\tilde{Z}(y')~~
\rm{with}~~y'=\exp \left[ \frac{4\pi^2}{\ln y}\right].
\label{math_02}
\end{eqnarray}
In the low frequency region $y=\exp(-x)\sim 1$ so that
$y'=\exp(-4\pi^2/x)\sim 0$ and $\tilde{Z}(y')\sim 1$. Low
frequency approximations of the free energy (\ref{equa11}) and of
the internal energy (\ref{equa21}) follow. There are similar
formulas associated with the other rational points of the circle
as shown below.

To get the leading term in $p(n)$ one uses the Cauchy formula
%
%
%
\begin{equation}
p(n)=\frac{1}{2i\pi}\oint \frac{\tilde{Z}(y)}{y^{n+1}}dy,
\label{math_03}
\end{equation}
where $\oint$ means an arbitrary closed loop encircling the
origin.

Substituting (\ref{math_02}) in (\ref{math_03}) with
$\tilde{Z}(y')=1$ one can obtain
\begin{equation}
p(n)=\frac{1}{2\pi\sqrt{2}}\frac{d}{dn}(\frac{\exp(K\lambda_n)}{\lambda_n})
\rm{with}~~\lambda_n=\sqrt{n-\frac{1}{24}}~~\rm{and}~~K=\pi\sqrt{\frac
{2}{3}}. \label{math04}
\end{equation}
This includes (\ref{equa7}) in the limit $n \rightarrow \infty$
but is much more accurate.
\subsection{Farey contributions and Ford circles}
\label{Ford}
From now on we extend the domain of definition of $\tilde{Z}(y)$
to the complex plane and we introduce the new variable $\tau$ and
Dedekind eta function $\eta(\tau)$ as defined in
(\ref{equa14bis})-(\ref{equa16}).

It can be shown that $\eta(\tau)$ is a modular form of degree
$-1/2$ on the full modular group. It acts on the generators of
such a group through the relations \cite{Apostol} \footnote{ For
more general modular transformations, we have
\begin{eqnarray}
\eta(\frac{p\tau+p'}{q\tau+q'})=\epsilon
(p,p',q,q')\sqrt{\frac{q\tau+q'}{i}}\eta(\tau)\nonumber,
\label{noteM_1}
\end{eqnarray}
with $\epsilon$ a $24th$ root of unity related to Dedekind sums as
defined in (\ref{math_14}). See \cite{Rademacher}, p. 160.

}
\begin{equation}
\eta(\tau+1)=\exp(i\pi/12)\eta(\tau);~~\eta(-1/\tau)=(\eta/i)^{1/2}\eta(\tau).
\label{math_3}
\end{equation}
To express the partition function one uses the Cauchy formula
\footnote{ Let $f(z)$ be an holomorphic function of the complex
variable $z$. The Cauchy formula for the derivatives is as follows
\begin{eqnarray}
f^{n}(a)=\frac{n!}{2i\pi}\oint \frac{f(z)}{(z-a)^{n+1}}dz
\nonumber \label{noteM_2}
\end{eqnarray}
where $\oint$ means a closed contour encircling the pole. It is
applied in (\ref{math_4}) to the partition function $f \equiv Z$
with $a=0$ so that $\frac{f^n(a)}{n!}=p(n)$ }
\begin{equation}
p(n)=\frac{1}{2i\pi}\oint
\frac{\tilde{Z}(y)}{y^{n+1}}dy=\int_{\tau_0}^{\tau_0+1}\tilde{Z}\left[
\exp(2i\pi \tau)\right] \exp(-2i\pi\tau n)d\tau. \label{math_4}
\end{equation}
In the third term above this corresponds to a path of unit length
starting at an arbitrary point in $\cal H$.

The choice of the integration path comes along in a natural way by
using the connexion of $\tilde{Z}(y)$ to the modular group. Let us
observe that the set of images of the line $\tau=X+i$, $X$ real,
under all modular transformations
\begin{equation}
\tau'=\frac{p\tau+p'}{q\tau+q'},~~\rm{with}~p,p',q,q'~\rm{integers}~\rm{and}~|pq'-q'p|=1,
\label{math_5}
\end{equation}
can be written as
\begin{equation}
\left| \tau-(\frac{p}{q}+\frac{i}{2q^2})\right|=\frac{1}{2q^2}.
\label{math_6}
\end{equation}
Equation \ref{math_6} defines circles $C(p,q)$ centered at points
$\tau=\frac{p}{q}+\frac{i}{2q^2}$ with radius $1/2q^2$. They are
named after L.R. Ford who first studied their properties in 1938
\cite{Rademacher}. Ford circles are easily generated by using the
ordered Farey sequence
\begin{equation}
\frac{0}{1}<\cdots<\frac{p_1}{q_1}<\frac{p_1+p_2}{q_1+q_2}<\frac{p_2}{q_2}<\cdots<\frac{1}{1}.
\label{math_7}
\end{equation}
%


%
To each $\frac{p}{q}$ belongs a Ford circle in the upper half
plane, which is tangent to the real axis at $\tau=\frac{p}{q}$. It
can be observed that Ford circles never intersect. They are
tangent to each other if and only if they belong to fractions
which are adjacent in some Farey sequence.

If $\frac{p_1}{q_1}<\frac{p}{q}<\frac{p_2}{q_2}$ are three
adjacent fractions in a Farey sequence then $C(p,q)$ touches
$C(p_1,q_1)$ and $C(p_2,q_2)$ respectively at the points
\begin{equation}
\tau_{pq}^L=\frac{p}{q}+\zeta_{pq}^L~~\rm{and}~~\tau_{pq}^R=\frac{p}{q}+\zeta_{pq}^R,
\label{math_8}
\end{equation}
where
\begin{equation}
\zeta_{pq}^L=-\frac{q_1}{q(q^2+q_1^2)}+\frac{i}{q^2+q_1^2}~~\rm{and}~~
\zeta_{pq}^R=\frac{q_2}{q(q^2+q_2^2)}+\frac{i}{q^2+q_2^2}.
\label{math_9}
\end{equation}
%
%


\begin{figure}[htbp]
\centering{\resizebox{7cm}{!}{\includegraphics{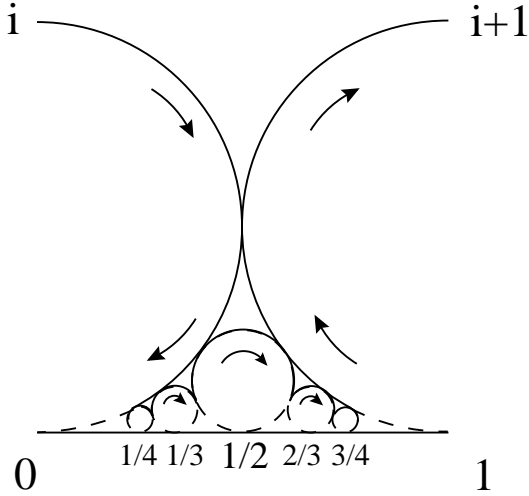}}}
\caption{Rademacher's path of integration.}
\end{figure}

In Rademacher's approach (which improves Ramanujan's one) the unit
length path on $\cal H$ is choosen so as to go along Ford circles
\begin{equation}
p(n)=\sum _{0\le p\le q\le
N}^{(p,q)=1}\int_{\gamma_{pq}}\tilde{Z}\left[\exp(2i\pi\tau)\right]
\exp(-2i\pi\tau n)d\tau, \label{math_10}
\end{equation}
where $\gamma_{pq}$ is the upper arc on a Ford circle which
connects points of tangency at $\tau_{pq}^L$ and $\tau_{pq}^R$.

Each Ford circle C(p,q) is parametrized by the expression
$\tau=\frac{p}{q}+\zeta$, where the variable $\zeta$ runs on an
arc of the circle $|\zeta_-\frac{i}{2q^2}|=\frac{1}{2q^2}$. If one
uses $z$ such that $\zeta=\frac{iz}{q^2}$, a Ford circle is mapped
onto the circle $|z-\frac{1}{2}|=\frac{1}{2}$ and (\ref{math_10})
transforms as
\begin{eqnarray}
&&p(n)=\sum _{0\le p\le q\le N}^{(p,q)=1} \{
\frac{i}{q^2}\exp(-2i\pi n\frac{p}{q})\nonumber \\ &&\times
\int_{z_{pq}^{\rm{L}}}^{z_{pq}^{\rm{R}}} \tilde{Z}\left[\exp (2\pi
i\frac{p}{q}-\frac{2\pi z}{q^2}) \right]\exp(\frac{2\pi
nz}{q^2})dz \}, \label{math_11}
\end{eqnarray}
where $z_{pq}^{\rm{L}}$ and $z_{pq}^{\rm{R}}$ follows from
(\ref{math_8}).

\subsection{Farey contributions and Dedekind sums}
\label{Dedekind} To compute (\ref{math_11}) one uses the
transformation formula $(\ref{equa14bis})$. After some
manipulations and using $z/q$ instead of $z$(see Ref.
\cite{Rademacher}, p. 269), one gets the formula which generalizes
(\ref{math_02})
\begin{equation}
\tilde{Z}(y)=\omega_{pq}\left(\frac{z}{q}\right)^{1/2}\exp(\frac{\pi}{12z}-
\frac{\pi z}{12 q^2})\tilde{Z}(y'), \label{math_12}
\end{equation}
with
\begin{equation}
y=\exp(\frac{2i\pi p}{q}-\frac{2\pi z}{q^2}),~y'=\exp(\frac{2i\pi
p'}{q}-\frac{2\pi}{z})\rm{and}~~pp'=-1(mod~q). \label{math_13}
\end{equation}
The so-called Dedekind sums $s(p,q)$ are introduced by
\begin{equation}
\omega_{pq}=\exp \left (i\pi s(p,q)\right ) ~~\rm{with} ~~s(p,q)=
\sum_{l=1}^q\left(
\frac{l}{q}\right)\left(\frac{pl}{q}-\left[\frac{pl}{q} \right ]
\right), \label{math_14}
\end{equation}
where $\left[~\right]$ in (\ref{math_14}) denotes the integer
part.

For the calculation of $p(n)$ one uses an approximation similar to
the one used in (\ref{math_02}). If $z$ is a small positive real
number, then $y$ is near $\exp(2i\pi\frac{p}{q})$, the modulus at
that point $|y'|=\exp(-\frac{2\pi}{z}) \sim 0$ and
$\tilde{Z}(y')\sim 1$.

As a result (\ref{math_11}) can be readily integrated and the
final result is
\begin{equation}
p(n)=\frac{1}{\pi \sqrt{2}}\sum_{q\ge1}\sqrt{q}A_q(n)\frac{d}{dn}
\left(  \frac{\sinh(K_q\lambda_n)}{\lambda_n}\right),
\label{math_15}
\end{equation}
\begin{eqnarray}
&&\rm{with}~K_q=\frac{\pi}{q}\sqrt{\frac{2}{3}},~~\lambda_n=\sqrt{n-\frac{1}{24}},\nonumber
\\
&&\rm{and}~A_q(n)=\sum_{p ~\rm{mod}(q)}\omega_{pq}\exp(-2i\pi
n\frac{p}{q}). \label{math_16}
\end{eqnarray}

\section{Conclusion}

We have found that quantum $1/f$ noise may be a property of
photons or phonons in thermal equilibrium. The theory connects
elliptic modular functions and quantum statistical mechanics as in
superstring theory, but with physical relevance in the macroscopic
realm of infrared divergences of solid state physics. Main results
are an enhanced energy per mode at low frequency (one gets
$\frac{\tilde{E}}{kT}\simeq \frac{\pi^2 k T}{6 h \nu}$ instead of
the Planck result $\tilde{E} \simeq kT$) and an enhanced
integrated radiation, photon density and phonon specific heat
(with a factor of $\zeta(3) \simeq 1.20$). Low frequency
fluctuations are reduced and one obtains fractional energy
fluctuations with a $1/\nu$ spectrum in contrast to the random
walk $1/\nu^2$ of the standard theory.

Main steps of the general mathematical theory of $p(n)$ is
reminded in the last section. It is based on the fact that there
are rational singularities on the unit circle in the partition
function. The fundamental mode in the Farey decomposition leads to
a satisfactory account of the infrared part of the spectrum and
remains correct in the high frequency region up to $x=x_1=4\pi^2$.
The physical meaning of higher modes has still to be understood.

Farey series and Ford circles were used recently in the different
context of $1/f$ noise in phase locked loops \cite{PRL02}. In this
last case they connect to the arithmetical functions found in
prime number theory.

\newpage
\section*{Acknowledgements} The authors wish to thank Professor
Handel for his generous invitation to the quantum $1/f$ noise
meeting in St. Louis. They also thank Professors Handel and Van
Vliet for their help in improving the content of the manuscript.

\end{document}